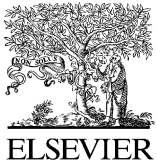

# FLUKA as a new high energy cosmic ray generator

Giuseppe Battistoni[a], Annarita Margiotta[b*], Silvia Muraro[a], Maximiliano Sioli[b]

[a] *INFN, Sezione di Milano, Via Celoria 16, I-20133, Milano, Italy*

[b] *Dipartimento di Fisica dell' Università di Bologna and INFN, Sezione di Bologna,*

*V.le Berti Pichat 6/2, I-40127, Bologna, Italy*

**Abstract**

FLUKA is a multipurpose Monte Carlo code, which can transport particles over a wide range of energies in user-defined geometries. Here we present a new FLUKA library, which allows the interaction and propagation of high energy cosmic rays in the Earth atmosphere and the transport of high energy muons in underground/underwater environments. © 2010 Elsevier Science. All rights reserved

## 1. Introduction

Underwater neutrino telescopes are optimized to detect upward going muons induced by high energy neutrinos. Despite the huge shielding effect of water, the main signal is due to downward going atmospheric muons. They result mainly from the decay of pions and kaons produced during the development of the extensive air showers initiated by high energy cosmic ray particles interacting with the atmospheric nuclei. The atmospheric muons are a background for neutrino detection, but they are useful also to check the detector response.

In both cases a reliable simulation of their abundance is required. Several Monte Carlo codes have been developed in this last few years to this purpose. At present the most diffused package for cosmic ray induced shower simulation is CORSIKA [1]. It allows the choice among different models for the description of extensive air shower formation and propagation and for the primary cosmic ray composition. An alternative approach is represented by the use of parameterized formulae, as is done in MUPAGE case [2] which provides a quick, but not flexible, simulation of underwater muon flux.

In this paper an application of the FLUKA package to the study of underwater muon flux is presented.

---
* Corresponding author. Tel.: +39-051-2095226; fax: +39-051-2095269; e-mail: margiotta@bo.infn.it.

## 2. The FLUKA code

The FLUKA package [3] is a general purpose Monte Carlo code used for the simulation and transport of particles developed since 1988 mainly by INFN and CERN researchers. It is used in many fields of physics for fundamental research and for a wide range of applications: accelerator physics, calorimetry, shielding design, radioprotection, dosimetry and, recently, hadron therapy. Since a few years it has been used also in space and cosmic ray physics.

These applications require precise and updated physics models. The FLUKA code is based on a theoretical microscopic approach with the advantage, with respect to parameterized models, of preserving correlations and of being predictive in the regions where experimental data are not available. The model parameters are fixed for all projectile-target combinations and energies and cannot be modified by the user. The theoretical models are continuously benchmarked against newly available data.

A combinatorial geometry package allows the definition of complex geometries that can be visualized through dedicated graphical tools. This allows simulating 3-D particle propagation and, as far as cosmic rays are concerned, to study showers with any angle and direction with respect to the Earth. It also takes into account the effect of different atmospheric compositions and magnetic field configurations. The same code allows the definition of cosmic ray detectors and their surrounding materials. The general FLUKA code is available on the website *http://www.fluka.org*. The development and maintenance are performed under an INFN–CERN agreement.

In this framework a new generator for high energy cosmic rays has been developed to extend the existing FLUKA cosmic ray library up to the TeV region.

## 3. muTeV generator

muTeV is a FLUKA based generator for TeV muons, validated and benchmarked along the years with the latest experimental results [4,5]. Moreover, it is a self-consistent generator that handles all the simulation steps in a unique framework, from the primary interaction in the atmosphere to the sampling at the detector level, going through the shower development in the atmosphere and the particle transport in the overburden.

It is mainly dedicated to the physics of high energy muons detected underground or underwater. It is optimized for TeV cosmic ray muons thanks to many biasing solutions adopted in different phases of the simulation to speed up the production chain. Moreover the package is flexible enough to include different underground / underwater sites. Presently, a version for an underground site (Laboratori Nazionali del Gran Sasso) and another one for underwater sites are available.

In the following only the underwater case will be described. A detailed presentation of the underground site version, is available in [4].

The main ingredients of the muTeV generator are:
- the physics models: they are taken from the FLUKA general code. The modeling of hadronic interactions is presently performed using the DPMJET package [6].
- the geometry setup: using the combinatorial geometry package included into the general code, the atmosphere is modeled with a set of 100 concentric spherical shells. Concerning the overburden, a water spherical shell with external radius $R_{ext}$=6378.14 km (average radius of the Earth globe) and internal radius depending on the site depth is defined. Two deep underwater sites have been considered up to now: the ANTARES [7] site $R_{int}$=(6378.14 – 2.475) km, and one of the proposed sites for the future km$^3$ neutrino telescope in the Mediterranean Sea [8] – the Capo Passero site- with $R_{int}$ = (6378.14 – 3.5) km. Only a modification of the geometry file is required to adapt the package to different underwater depths. The sampling level of the shower axis is defined on a circle (*sampling circle*), whose orientation is perpendicular to the shower direction and centered in the centre of a virtual cylinder surrounding the instrumented region and defining the active volume of the neutrino telescope (the so called *can*). The radius $R_s$ of the *sampling circle* and the dimensions of the *can* depend on the detector size. In the ANTARES case $R_s$=500 m while the height and the radius of the cylinder are H= 600 m and $R_c$=240 m. For the Capo Passero site $R_s$ = 1000 m, while H=1000 m and $R_c$=500 m.

Once the primary direction is randomly extracted, a point is sampled on the *sampling circle*. The primary direction is then traced back to the upper atmosphere to locate the primary injection point. Finally, the particle is propagated through the atmosphere, the shower is developed and muons are followed through water to the surface of the *can* where their position, angles and energy are registered.

The effect of the Earth magnetic field is taken into account.

- the source beam: it is defined by the particles to be propagated throughout the geometry setup, their kinetic energy, injection point and direction. In the present version particles are sampled from a primary mass composition model described in [9], which can be changed by the user. For each primary nucleus and for each direction, which defines the amount of water to be crossed, the energy threshold required to produce at least one muon underwater with a predefined small probability value is computed. Particles are then followed in the atmosphere until their energy is lower than the threshold.

This set of information, supplied by the user through data cards, is necessary to start a FLUKA run. Each run contains a user-defined number of *stories*, corresponding to a detector livetime provided by FLUKA itself.

Fig. 1 shows the distribution of the vertexes of muons entering the active volume surface for a km$^3$ detector at a depth corresponding to the Capo Passero site.

A dedicated version of the muTeV library is in preparation to study coincidences between neutrino telescopes and surface arrays with the aim of defining the absolute orientation of the floating underwater detectors.

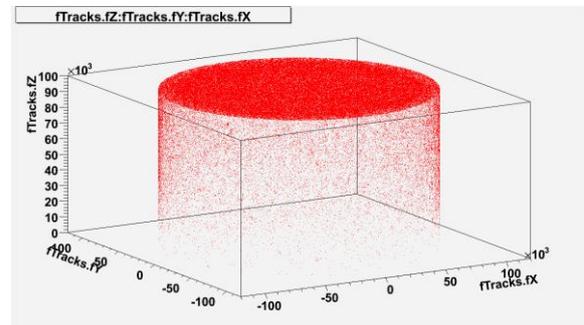

Fig. 1. Vertexes of particles entering the surface of the active volume of a km$^3$ detector at the Capo Passero site depth.

## 4. Conclusions

A new library for the generation of atmospheric muon flux underwater was developed in the framework of the FLUKA simulation package. At present two different geometry setups are already defined, but the application can be easily extended to other sites.

## Acknowledgments

This work has been carried out in the framework of the FLUKA Collaboration.